\documentclass[12pt,a4paper]{article}

\usepackage[margin=1in]{geometry}
\usepackage{setspace}
\doublespace                      
\usepackage{epsfig}
\usepackage{graphicx}
\usepackage{bm}
\usepackage{amsfonts}
\usepackage{amssymb}
\usepackage{amsmath}
\usepackage{amsthm}

\usepackage{cases}
\usepackage[justification=centering]{caption}
\usepackage[bookmarks=true,colorlinks=true,linktocpage=true,backref=true]{hyperref}
\usepackage[margin=1in]{geometry}
\usepackage{tensor}
\usepackage[table]{xcolor}
\usepackage{kbordermatrix}
\usepackage{wasysym}
\usepackage{wrapfig}
\usepackage[numbers]{natbib}
\usepackage{url}
\usepackage{doi}

\DeclareCaptionLabelFormat{figlabel}{\textbf{Fig. #2}}

\begin{document}
\fontfamily{cmr}
\fontseries{m}
\fontshape{n}
\selectfont

\thispagestyle{empty}

\begin{center}

{\Large {\bf A No-Go Theorem for Quantum Cosmologies \\
with Non-natural Hamiltonians}}

\vspace{0.5cm}

\normalsize

Christine C. Dantas

\vspace{0.5cm}

\footnotesize{Divis\~ao de Astrof\'{\i}sica \\
Instituto Nacional de Pesquisas Espaciais,\\
Av. dos Astronautas, 1758, Jardim da Granja, \\
CEP: 12227-010, S\~ao Jos\'e dos Campos, SP, Brazil \\ 
E-mail: {\tt christine.dantas@inpe.br}
}

\end{center}

\vspace{1.0cm}

\begin{quotation}

{\bf Abstract.} The Eisenhart–Duval lift (ED) geometrizes classical dynamics by embedding their trajectories into null geodesics of a higher-dimensional Lorentzian spacetime. However, such a construction requires a natural Hamiltonian, that is, quadratic in the canonical momenta. As a consequence, mini\-superspace cosmological models governed by non-natural Hamiltonians cannot admit an ED lift. Effective models in Loop Quantum Cosmology provide a concrete example: polymer-modified Hamiltonians become non-polynomial in the momenta and therefore fall outside the metric framework of the ED lift. We thus establish a kinematical no-go theorem: non-quadratic cosmological dynamics cannot be geometrized via ED constructions. Quantum-corrected bounce models therefore illustrate a structural limitation of metric geometrization within the ED framework.
A few qualitative comments on the possibility of Stäckel-lifting a single-degree-of-freedom non-natural Hamiltonian in quantum cosmology is also presented.

\end{quotation}

\vspace{2cm} 
\begin{center}
{\scriptsize \sl This essay received an Honorable Mention in the 2026 Essay Competition of the Gravity Research Foundation.}
\end{center}

\clearpage
\setcounter{page}{1}

\noindent {\bf Introduction.} 

$~$ 

In the late 1920s, Eisenhart \cite{Eis28} published a remarkable paper in which he proved that the equations of motion of classical, natural Hamiltonian systems (i.e., quadratic in the momenta), with $n$ degrees of freedom, could be reframed as the geodesic equations in a certain, ``lifted'' Lorentzian spacetime of $n+2$ dimensions or more. About half a century later, Eisenhart's  framework (unnoticed until recently) was rediscovered independently by Duval et al. and others (cf. \cite{Duv85,Duv91,Duv24}, and references therein). Hereon, we denote this construction as the {\sl Eisenhart–Duval lift} (ED lift; for a didactical exposition, cf. \cite{Car15}). Over the past years, this formalism has been refined to explore the symmetries of Galilei-invariant, classical and quantum-mechanical systems, in terms of properties of their lifted spacetimes, a procedure which is somewhat reminiscent of Kaluza-Klein theory (e.g., \cite{Ove97}, and references therein).

The ED lifted spacetimes are Lorentzian manifolds admitting a covariantly constant null vector field. Such geometries were first systematically analyzed by Brinkmann \cite{Bri25}, in his study of conformal mappings between Einstein spaces. In particular, he showed that any Lorentzian manifold possessing a parallel null vector field can locally be written in a canonical coordinate system, now known as Brinkmann coordinates, which makes the null structure manifest. In the ED construction, such a null vector field generates a free isometric action whose quotient defines a nonrelativistic spacetime equipped with a Bargmann structure (to be described in the next section, cf. \cite{Bar54,Duv85}). 
Indeed, the ED lift can be placed in a broader geometric framework, in which the conformal isometries of the Bargmann spacetime encode the Schrödinger group in special cases \cite{Duv24}. 

Recently, Cariglia et al.~\cite{Car18} (hereon, [Car+18]) extended the ED lift to the Friedmann--Lemaître--Robertson--Walker (FLRW) cosmological model~\cite{Weinberg1972,MTW1973}. In this setting, the cosmological dynamics can be mapped to a time-dependent mechanical system~\cite{Car16}, with the time-dependent harmonic oscillator providing a convenient conceptual model. The latter is naturally associated with the Ermakov--Milne--Pinney (EMP) equation~\cite{Ermakov1880English,Milne1930,Pinney1950}, which captures the dynamics of the corresponding time-dependent oscillator. In recent years, the EMP equation has been studied in both classical and quantum cosmology (see~\cite{Car18} and references therein). In particular, [Car+18] showed that the Friedmann equations can be recast within the ED lift framework, where the evolution of the lifted spacetime is encoded in an EMP-type equation.

In the present essay, we comment on the structural incompatibility between minisuperspace cosmological dynamics and the ED lift (e.g., \cite{Kiefer2012} for a systematic discussion of minisuperspace quantization in canonical quantum cosmology). We summarize this fact in terms of a general no-go theorem:  {\sl single-degree-of-freedom minisuperspace Hamiltonians (governing the evolution in relational time) that are  not quadratic in the canonical momentum do not admit an  ED lift}.  
This obstruction is purely kinematical and follows from the  geometric requirements underlying the lifted Bargmann structure. As a consequence, any effective cosmological model whose Hamiltonian 
departs from quadratic momentum dependence lies outside the ED framework.  {The  possibility of more generally lifts for such Hamiltonians, however, are taken into consideration as preliminary comments in the last section before the conclusions of this work.}

It is important to clarify a few points. By construction,  any non-quadratic Hamiltonian cannot be ED-lifted, and this is just a corollary of the ED geometrization, which is certainly not a novel observation. As such, this is not the specific claim of the present no-go theorem. Instead, this well-known obstruction (namely, non-quadratic Hamiltonian cannot be ED-lifted) is highlighted for the case of quantum cosmological models with such Hamiltonians, as in minisuperspace models occurring in Loop Quantum Cosmology (LQC) \cite{Boj08}. Our aim is to provide a compact statement regarding this latter fact in terms of a no-go theorem for such quantum cosmological scenarios and to comment on its general consequences.

LQC relies on quantization techniques of Loop Quantum Gravity (LQG) \cite{Rovelli2004, Thiemann2007,Kiefer2012,RovelliVidotto2015} to symmetry-reduced geometries, resulting in a quantization of cosmological degrees of freedom and modified effective dynamics at Planckian densities. These so-called polymer-modified effective cosmologies in LQC provide an illustration of our no-go theorem, as their Hamiltonian become non-polynomial in the canonical momenta  \cite{AshtekarSingh2011,Ash06,Ash06b}. The main interest here is that effective equations in LQC could result in non-singular cosmological bounces (NSCBs) \cite{Ash06,Ash06b}, which has lead to an active area of research as well as important critical analyses (e.g. \cite{Singh_2009,Boj25}). 

NSCBs necessarily evade the classical singularity theorems and the standard Friedmann dynamics of general relativity. As reviewed in \cite{BrandenbergerPeter2017}, NSCB in general require at least one of the following: (i) violation of the null energy condition, (ii) modification of the gravitational dynamics, or (iii) departure from canonical quadratic Hamiltonian structure. Our no-go theorem concerns exclusively the third possibility. Note that the ED framework has been used as an interpretative bridge between the original dynamics and the lifted one. Many cosmological bounce models implicitly assume geometrizability, but our theorem cuts that bridge, as NSCBs evade geometric reinterpretation within the ED paradigm.
This leads to concrete consequences, to be discussed in the final section.

$~$ 

\noindent {\bf Ingredients for the Eisenhart–Duval Lift.}

$~$ 

We briefly outline the background material leading to the ED lift in order to subsequently establish our no-go theorem with the minimum underlying prerequisites.

First, a {\sl Brinkmann spacetime} is a Lorentzian manifold $(M,g)$ admitting a covariantly constant null vector field $k$, i.e.
\begin{equation}
\nabla_\mu k^\nu = 0,
\qquad
k^\mu k_\mu = 0.
\label{eq:cov_null}
\end{equation}
These conditions define a parallel null vector field and therefore endow the spacetime with a preferred null direction. A classical result of Brinkmann states that any spacetime admitting such a vector field can be locally written in a canonical coordinate system. In $(n+2)$ dimensions, the metric takes the form
\begin{equation}
ds^2
=
g_{ij}(u,x)\, dx^i dx^j
+ 2\, du\, dv
+ H(u,x)\, du^2,
\label{eq:Brinkmann}
\end{equation}
where $x=(x^1, \dots ,x^n)$ are transverse spatial coordinates, $u$ is a null coordinate, and $v$ parametrizes the null direction. In these coordinates, the parallel null vector field is manifestly
$
k = \partial_v .
$

A {\sl Bargmann structure} is a Lorentzian manifold $(M^{n+2},g)$ equipped with (i) a covariantly constant null vector field $k$, and (ii) a free null action generated by $k$. In adapted coordinates $(x^i,t,\sigma)$, the metric can be written as
\begin{equation}
ds^2
=
g_{ij}(x,t)\, dx^i dx^j
+ 2\, dt\, d\sigma
- 2 V(x,t)\, dt^2,
\label{eq:adapted}
\end{equation}
with $k = \partial_{\sigma}$. This is precisely the metric appearing in the ED lift of classical mechanics.
One may form the quotient:
\begin{equation}
\mathcal{N} = M / \mathrm{span}(k), \label{eq:N}
\end{equation}
\noindent which inherits a Newton--Cartan structure \cite{Duv85,Trautman1965NewtonCartan}.
The quotient along the integral curves of $k$ yields the underlying Newtonian spacetime, while the conserved momentum conjugate to the null direction is interpreted as mass. This geometric framework, later systematized by Duval and collaborators, provides the natural setting for the ED lift and its relation to the centrally extended Galilei (Bargmann) symmetry.

Let $(q,p)$ denote canonical coordinates on a two-dimensional phase space equipped with canonical symplectic form, $\omega = dq \wedge dp$. Consider a model whose dynamics is generated by a Hamiltonian $H(q,p)$.
A necessary and sufficient condition for the existence of a genuine ED lift of the dynamics, within the standard ED/Bargmann construction, is that the Hamiltonian be of {\sl natural form}, i.e.
\begin{equation}
H(q,p) = \frac{1}{2} g^{ij}(q)\, p_ip_j + V(q), \label{eq:natural}
\end{equation}
with quadratic kinetic term (i.e., polynomial degree 2 in canonical momenta), also containing dependence on configuration-space metric, and canonical symplectic structure. 

The mechanism of the ED lift is the following. Given a natural Hamiltonian (Eq. \ref{eq:natural}), one introduces an extended configuration space by adding auxiliary coordinates and constructs a Lorentzian metric whose geodesics reproduce the original dynamics after null reduction. The kinetic term determines the spatial components of the metric, while the potential is encoded in specific off-diagonal metric components coupling the physical coordinates to the additional variables. The resulting geodesic Lagrangian is quadratic in velocities, and therefore its (geometric) Hamiltonian is quadratic in all canonical momenta. 

$~$ 

\noindent {\bf A Structural No--Go Theorem for Eisenhart--Duval Lifts.}

$~$ 

We consider minisuperspace models of LQC in which a scalar field serves as an internal clock. After solving the Hamiltonian constraint for the conjugate momentum of the clock variable, the system reduces to a single-degree-of-freedom Hamiltonian governing evolution in relational time. It is this deparametrized mechanical system that is shown here to be obstructed to an ED lift.  Schematically, at the fundamental canonical gravity level, given a constraint, $\mathcal{C}(q,p) = 0$, there is no true Hamiltonian generating time evolution, as time is gauge.
The ED lift does not directly apply here, because ED requires $H(q,p)$ as a true Hamiltonian generating evolution. By choosing an internal clock (e.g., scalar field, $\phi$) and solving the constraint for its conjugate momentum ($\mathcal{C} = 0 \Rightarrow p_\phi = \pm H_{\text{true}}(a, p_a)$), this deparametrization yields an ordinary Hamiltonian, $H_{\text{true}}(a,p_a)$, governing evolution in the relational time $\phi$, so the ED lift applies in this case.

We now formulate the structural obstruction underlying the absence of a ED lift for single-degree-of-freedom, non-quadratic minisuperspace dynamics in canonical quantum cosmology, with a relevant example being effective polymerized models in LQC. It will be made clear that the obstruction is algebraic before it is geometric.

$~$ 

\underline{\sc Theorem (Structural No-Go)}

{\sl 
Let the Hamiltoninan $H(p,q)$ fail to be polynomial of degree two in canonical momenta. Then no Lorentzian metric on a finite-dimensional manifold reproduces its Hamiltonian flow via null geodesics in the Eisenhart–Duval construction.
In particular, the ED lift is impossible for single-degree-of-freedom, non-quadratic, deparametrized minisuperspace Hamiltonians in canonical quantum cosmology. }

\underline{{\sc Proof:}}
The obstruction follows from the defining structure of the ED lift. In this construction, the dynamics of a mechanical system are embedded into null geodesics of a Lorentzian metric in an extended manifold. The associated geodesic Hamiltonian is necessarily quadratic in all canonical momenta, being of the form:
\begin{equation}
H_{\rm geo}= \frac{1}{2} G^{AB}(x) P_AP_B. \label{eq:Hgeo}
\end{equation}
\noindent Since projection onto the original system preserves the polynomial degree in momenta, only Hamiltonians that are quadratic in canonical momenta (i.e. of natural form) can arise from such a metric contraction. A Hamiltonian containing non-polynomial or higher-order momentum dependence therefore cannot be generated by any finite-dimensional Lorentzian metric via the ED construction. The obstruction is structural and independent of dimension. 

In effective LQC, the Hamiltonian typically takes the schematic form
\begin{equation}
H_{\mathrm{LQC}}(q,p)
=
A(q)\, \frac{\sin^2(\lambda p)}{\lambda^2} + V(q), \label{eq:A}
\end{equation}
\noindent which is non-polynomial and periodic in $p$. Such Hamiltonians are not quadratic in momenta and therefore violate the necessary condition established above. The obstruction is global and not removed by perturbative expansion around small polymer parameter. Hence no ED lift exists for polymerized cosmological dynamics. \hfill $\square$

Note that in the trivial small-$\lambda p$ limit the classical quadratic term is recovered, but this regime is not of physical interest for the polymer-modified theory. One might object that the non-polynomial term is analytic and expandable, so one could attempt a local quadratic approximation to restore the conditions for an ED lift. However, this is not possible as well, since the obstruction is global and structural, and therefore a local perturbative quadratic approximation  does not restore exact lift. Nor can the non-polynomial structure be absorbed into a redefinition of the metric. Any finite-dimensional phase-space extension introducing auxiliary variables still leads to a geodesic Hamiltonian quadratic in all canonical momenta, and thus cannot reproduce the original non-polynomial dependence without altering the physical content of the theory.

\clearpage

\noindent {\bf A possible lift via a generalized Ermakov–Milne–Pinney equation?}

$~$ 

As mentioned in the Introduction, [Car+18] \cite{Car18} extended the ED lift to FLRW cosmology by associating the lifted dynamics with an EMP equation. One might therefore wonder whether a modified Friedmann equation arising in effective quantum cosmology (e.g. in LQC) could similarly be rewritten as a generalized EMP equation, thereby evading the no-go result established above.

At the level of differential equations, this appears conceivable. The EMP equation,
\begin{equation}
\ddot a = -\omega^2(t)\, a + \frac{\lambda}{a^3},
\end{equation}
is a second-order nonlinear ODE. Likewise, effective LQC dynamics leads to a modified Raychaudhuri equation for the scale factor $a(t)$ derived from the modified Friedmann equation. Formally, such an equation may be expressible in the schematic form
\begin{equation}
\ddot a = -\omega^2(t)\, a + g(a),
\end{equation}
with $g(a)$ encoding nonlinear corrections (although such an identification may prove to be complicated).

However, this observation does not circumvent the obstruction. The existence of an ED lift is unrelated to the ability of rewriting a second-order ODE in EMP-like form. Rather, as we have seen, it is a statement about the underlying Hamiltonian required for embedding the dynamical trajectories into null geodesics of a higher-dimensional Lorentzian spacetime. Schematically,
\begin{equation}
\mathrm{Natural~Hamiltonian}
\;\Longleftrightarrow\;
\mathrm{ED~lift~exists}
\;\Longrightarrow\;
\mathrm{EMP~equation},
\end{equation}
whereas the converse implication does not hold:
\begin{equation}
\mathrm{Modified~EMP}
\;\not\Longrightarrow\;
\mathrm{ED~lift~exists}.
\end{equation}

Thus, even if one engineers a higher-dimensional metric reproducing a generalized EMP equation at the level of differential dynamics, such a construction would not constitute a genuine ED lift of the quantum-corrected cosmological Hamiltonian. An EMP-type rewriting encodes only the autonomy of the second-order ODE and does not restore the natural Hamiltonian structure required for an ED lift.

$~$

\noindent {\bf A possible geometrization via Stäckel lifts? }

$~$

We have described the absence of ED lifts for non-quadratic Hamiltonians, encompassing the important case of minisuperspace LQC models, as a structural no-go theorem. In this section, we briefly comment on other more general geometrical lifts. They can be seen as a generalization of the simpler, geodesic ED lift, in which {\sl the higher-dimensional metric also depends on the velocities or on the momenta}, which is not case of the lifted metric in Eq. \ref{eq:adapted}, depending only on the spatial/null coordinates. Such generalized metrics were motivated, for instance, in quantum gravity scenarios in which there are departures from the local Poincaré structure of spacetime (\cite{Rel21,Rel24}, and references therein).

In such scenarios, one would consider the possibility that photons with different energies travel at different velocities in vacuum, instead of a constant velocity ($c$). The change in their velocities would result in a very small time delay when photons travel across the Universe. This scenario could be modeled through a metric which depends on, besides of the spacetime coordinates, the vectors on the tangent spacetime manifold, i.e., on the velocities. In this case the formalism can be developed in a {\sl Finsler space} \cite{Rund59}, a particular case of a {\sl Lagrangian} space where the fundamental metric function must be positively homogeneous \cite{Vac08}. Another approach is a metric depending on the cotangent vectors, i.e., on the momenta, besides the spacetime coordinates (namely, depending on all the phase-space variables), in which case the formalism is developed in the so-called {\sl Generalized Hamilton Spaces} (hereon, GHS; \cite{Rel24}), which could also be modeled as higher-dimensional lifted spaces. Note that, broadly, the usual classical models in physics consider {\sl Hamilton spaces} with simple, fixed energy setups, that is, with fixed rules for the propagation of particles. In GHS, energy and distance rules may change depending on the momentum of the particle.

Regarding how general a Hamiltonian system can be  (i.e., with non-polynomial momentum dependencies), in order that it can be introduced into an appropriately defined, higher-dimensional GHS (or, equivalently, in order to be recovered from such a space), there seems to be no general established procedure. For instance, in Ref. \cite{Rel24}, the authors propose a new method to obtain all geometrical properties of a GHS, once a generic metric is given. Their method is the inverse of the usual approach (which can be overwhelming in general cases) in which one starts directly from the construction of a Hamilton space and then obtains the associated metric geometry \cite{Miron01,Miron12}.

On the other hand, a generalization of the ED lift has been recently proposed in  Ref. \cite{Kub25}, based on a {\sl generalized Stäckel matrix}. 
The procedure requires a separable Hamiltonian system, defined on the cotangent bundle, $T^*M$, of a 
$n$-dimensional manifold, $M$. Hence, there is a $n \times n$  Stäckel matrix associated with this system.
One could then extend this matrix to a generalized Stäckel matrix, hence defining a $(n+1) \times (n+1)$ lifting matrix. The latter includes the original Stäckel matrix, as well as additional row(s) and column(s), in which the elements are chosen functions depending on a the additional pair of conjugate variables, 
$(q^{n+1},p_{n+1})$ (this procedure can be iterated with new pairs). Hence the original $n$-dimensional manifold is extended to: $\tilde{M}= M \times \mathbb{R}$, with $q^{n+1}$ denoting the coordinate on 
$\mathbb{R}$, and  $T^*\tilde{M}$ is the extended cotangent bundle. The extended symplectic form is simply: 
$\tilde{\omega} = \sum_{i = 1}^{n+1} dq^i \wedge dp_i$.  

The idea is then that such a procedure leads to geodesic or even non-geodesic lifts that generalize ED lifts. Geometrically, {\sl the resulting 
lifted-Stäckel Hamiltonian system corresponds to imposing a momentum-dependent metric on} $T^* \tilde{M}$.
In summary, such generalized Stäckel lifts on Hamilton spaces have been admitted under conditions that are more general than those considered in the present work, which assumes the simpler ED/Bargmann construction.
The above-mentioned approaches \cite{Rel24, Kub25} raise an interesting question of whether our no-go theorem can be evaded for the case of generalized Stäckel lifts on GHS, that is, if such a construction can be defined for minisuperspace Hamiltonians in LQC, to begin with.

We summarize some definitions and conditions given in \cite{Kub25}. 
It is well known \cite{Per90} that a totally separable, integrable Hamiltonian system, $(H_1,\dots,H_n)$, with all $H_j$ quadratic in the momenta, can be characterized using an invertible, $(n \times n)$ Stäckel matrix, in which its components are functions of $\mathbf{q}\equiv (q_1,\dots,q_n)$, and a Stäckel vector for the potential, in which its components are also functions of $\mathbf{q}$.  We require, however, that the Stäckel matrix be not only a function of $\mathbf{q}$, but also of the momenta, $\mathbf{p}$ (see also Remark 5 in Ref. \cite{Kub25}). 
Furthermore, Arnold, Kozlov and Neishtadt (AKN; \cite{Arn97}) generalized such a representation for the case of arbitrary, Hamiltonian functions, not necessarily quadratic in the momenta, which is the case we will consider here. Let $S^{(n)}(\mathbf{q},\mathbf{p})$ be such an $(n \times n)$ Stäckel matrix, then it is generically lifted to the following matrix:

\begin{equation}
\mathcal{L}^{(n+1)} (\mathbf{q},\mathbf{p}) \equiv
\left [ \begin{array}{ccc|c}
S_{11}(q^1,p_1)  & \dots & S_{1n}(q^1,p_1)  & S_{1,n+1}(q^1,p_1) \\
 \vdots & \ddots &  \vdots &  \vdots \\
S_{n1}(q^n,p_n)  & \dots & S_{nn}(q^n,p_n)  & S_{n,n+1}(q^n,p_n) \\
\hline
0 & \dots & 0 & S_{n+1,n+1}(q^{n+1},p_{n+1}) 
\end{array} \right ] .\label{eq:liftedS}
\end{equation}

Then the corresponding lifted-Stäckel Hamiltonian system on $T^* \tilde{M}$ associated with Eq. \ref{eq:liftedS} can be generically written as:

\begin{equation}
\left [ \begin{array}{c}
H_1     \\
\vdots  \\
H_{n+1}  
\end{array} \right ]
=
\left (
\mathcal{L}^{(n+1)} (\mathbf{q},\mathbf{p}) 
\right )^{-1}
\left [ \begin{array}{c}
f_{1}(q^1,p_1) \\
 \vdots \\
 f_{n+1}(q^{n+1},p_{n+1}) 
\end{array} \right ] .\label{eq:liftedH}
\end{equation}

Clearly, in the case of single-degree-of-freedom minisuperspace LQC Hamiltonians (which we take again Eq. \ref{eq:A} as a representative example),  one would start with a $1 \times 1$ Stäckel matrix that is itself dependent on the momentum $p_{1}$ (via the $\sin^2$ function), besides the dependence on a generic function $A(q^1)$.  We would have, then, some $2 \times 2$ lifted-Stäckel Hamiltonian system, associated with Eqs. \ref{eq:liftedS} and \ref{eq:liftedH}, for some appropriate function, $f_{2}(q^{2},p_{2})$, for the additional variables.

As already mentioned, the case at hand involves a Hamiltonian that depend on momenta not only through the Stäckel functions $f_k$ in Eq. \ref{eq:liftedH}, but also through the coefficients of the inverse of the momentum-dependent matrix, Eq. \ref{eq:liftedS}; this situation corresponds to a momentum-dependent metric on $T^* \tilde{M}$ \cite{Kub25}. A complete geometric knowledge and corresponding properties of such resulting GHS could be approached systematically, for instance, once a generic metric was given \cite{Rel24}. Such properties rely on certain definitions regarding a Hamiltonian function to be the squared geodesic distance in the momentum space, which can be proved to be autoparallel and satisfying a differential equation. Tangent and cotangent spaces of $T^* \tilde{M}$ split into horizontal and vertical subspaces, where the movements along the base and through the fiber of the bundle geometry take place, respectively.

In that regard, the importance of autoparallel Hamiltonians is central. If a Hamiltonian
is an homogeneous function in momenta, then it is always autoparallel \cite{Bar15}, and if a given Hamiltonian is autoparallel, any function of it will be also autoparallel \cite{Rel24}. In the present case, due to the sinusoidal dependence on the momentum in the Stäckel matrix, the lifted Hamiltonian will not be a  homogeneous function in momenta ($\sin^2(\lambda p) \neq \lambda^r \sin^2(p)$, for any $r \in \mathbb{R}$, $\lambda \neq 0 \in \mathbb{R}$; cf. Eq. \ref{eq:A}). It is a general expectation that Planck-scale modified dispersion relations cannot be expressed in homogeneous Hamiltonians, and the resulting equations of motion will not be autoparallels of the geometry \cite{Bar15}, meaning, they will contain a force-like term. Analogously, if indeed single-degree-of-freedom minisuperspace LQC Hamiltonians can be generally Stäckel-lifted in the manner here outlined, it would be natural to expect similar non-geodesic flows. 

In Ref. \cite{Kub25}, several examples are given for non-geodesic flows leading to higher-order integrable systems. However, in the examples given, the dependence on momenta, introduced through the Stäckel functions $f_k$, {\sl are associated with potentials terms only}. If LQC Hamiltonians can be generally Stäckel-lifted, then still remains the task of showing that the resulting lifted space can be submitted to a proper geometric analysis, such as those given in \cite{Rel24, Bar15}. This matter deserves further investigation and it is left for future work.

\clearpage

\noindent {\bf Discussion and Conclusion.}

$~$

The Eisenhart–Duval (ED) lift relies essentially on the quadratic dependence of the Hamiltonian on the momenta. In the present essay, we elevate this seemly innocuous observation to a no-go theorem applicable to non-quadratic effective Hamiltonians in canonical quantum cosmology: their quantum-corrected models cannot be ED-lifted. Concretely, in LQC the ED lift is applied not to the constrained cosmological system itself, but to the deparametrized minisuperspace Hamiltonian obtained after choosing an internal clock.

Our statement highlights the failure of the ED lift as categorical, since it signals the breakdown of the lifted metric framework itself rather than an incomplete implementation of it.  As a result, quantum-corrected dynamics cannot be analyzed using the geometric machinery associated with ED lifted spaces. The obstruction is structural and precedes any discussion of dynamical symmetry. For instance, without an ED lift, the conformal geometric interpretation underlying Schrödinger symmetry is unavailable \cite{Duv24}. 

One may construct an effective spacetime whose scale factor satisfies a generalized EMP equation. However, such a construction does not arise from a Bargmann structure encoding the underlying Hamiltonian and therefore lacks the structural meaning of the ED lift. The failure is not dynamical but structural, as a quantum-corrected cosmology may mimic the form of a lifted system, but it cannot arise from one.

Importantly, our no-go theorem is structural in nature.  It applies to any single-degree-of-freedom minisuperspace model whose effective Hamiltonian departs from the natural quadratic dependence on the canonical momenta, independently of the microscopic origin of the effective dynamics. Thus, if a modified-gravity or exotic-matter bounce scenario is reduced to such a non-natural Hamiltonian framework, the obstruction to an ED lift persists. Our result therefore delineates a structural domain of applicability rather than excluding bounce scenarios on dynamical grounds.

The absence of an ED lift has concrete consequences. When such a lift exists, the dynamics can be embedded into a Lorentzian Bargmann spacetime, allowing one to interpret trajectories as null geodesics, conserved quantities as Killing symmetries, and stability properties in terms of curvature. These geometric tools provide powerful structural insights and often reveal hidden integrability properties. In polymerized cosmology, however, the non-quadratic Hamiltonian obstructs this embedding. As a result, quantum-corrected dynamics cannot be analyzed using the geometric machinery associated with Bargmann structures, and any effective geometric interpretation must depart from a lifted Lorentzian framework altogether.

A rigorous analysis on the  possibility of generally Stäckel-lifted, single-degree-of-freedom, non-natural minisuperspace LQC Hamiltonians, goes beyond the scope of the present work. We provided a few qualitative comments on the conditions required for the admission of such generalized lifts. These are important to be understood in detail and could potentially evade the simpler no-go theorem presented here.

$~$ \\

{\footnotesize  {\sl \underline{Acknowledgements}.}

CCD thanks the referee of the {\it Int. J. Mod. Phys. D} for the constructive comments, which have significantly improved this work. CCD thanks Brazilian Ministry of Science, Technology and Innovation, and
the Brazilian Space Agency (AEB), which supported the present work under PO 20VB.0009.


\clearpage

\bibliographystyle{iopart-num} 
\bibliography{CCDANTAS}

\end{document}